\documentclass[twocolumn,showpacs,preprintnumbers,amsmath,amssymb,superscriptaddress,floatfix]{revtex4}

\usepackage{graphicx}
\usepackage{dcolumn}
\usepackage{bm}

\begin{document}


\title{Experimental evidence of paired hole states in model high-$T_c$ compounds}%

\author{A.~Rusydi}
\affiliation{National Synchrotron Light Source, Brookhaven National
  Laboratory, Upton, NY, 11973-5000, USA} 
\affiliation{Materials Science Centre, University of Groningen, 9747 AG Groningen, The Netherlands}
\affiliation{Institut f\"{u}r Angewandte Physik,
Univerist\"{a}t Hamburg, Jungiusstra$\ss$e 11, D-20355 Hamburg, Germany}

\author{P.~Abbamonte}
\affiliation{National Synchrotron Light Source, Brookhaven National
  Laboratory, Upton, NY, 11973-5000, USA}
\affiliation{Physics Department and Frederick Seitz Materials Research
  Laboratory, University of Illinois, Urbana, IL, 61801} 

\author{M.~Berciu}
  \affiliation{Department of Physics and Astronomy, University of
  British Columbia, Vancouver, B.C., V6T-1Z1, Canada} 

\author{S.~Smadici}
\affiliation{National Synchrotron Light Source, Brookhaven National
  Laboratory, Upton, NY, 11973-5000, USA}
\affiliation{Physics Department and Frederick Seitz Materials Research
  Laboratory, University of Illinois, Urbana, IL, 61801} 
  
\author{H.~Eisaki}
\affiliation{Nanoelectronics Research Institute, AIST, 1-1-1 Central
  2, Umezono, Tsukuba, Ibaraki, 305-8568, Japan} 

\author{Y.~Fujimaki}
\affiliation{Department of Superconductivity, University of Tokyo, Bunkyo-ku, Tokyo 113, Japan}

\author{S.~Uchida}
\affiliation{Department of Superconductivity, University of Tokyo, Bunkyo-ku, Tokyo 113, Japan}

\author{M.~R\"{u}bhausen}
\affiliation{Institut f\"{u}r Angewandte Physik,
Univerist\"{a}t Hamburg, Jungiusstra$\ss$e 11, D-20355 Hamburg, Germany}

\author{G.~A. Sawatzky}
\affiliation{Department of Physics and Astronomy, University of
  British Columbia, Vancouver, B.C., V6T-1Z1, Canada}

\date{\today}

\begin{abstract}

The distribution of holes in Sr$_{14-x}$Ca$_x$Cu$_{24}$O$_{41}$ (SCCO) is revisited 
with semi-emperical reanalysis of the x-ray absorption (XAS) data and exact-diagonalized 
cluster calculations. A new interpretation of the XAS data leads to much larger 
ladder hole densities than previously suggested. These new hole densities 
lead to a simple interpretation of the hole crystal (HC) recently 
reported with 1/3 and 1/5 wave vectors along the ladder. 
Our interpretation is consistent with paired holes in the rung of the ladders. 
Exact diagonalization results for a minimal model of the doped 
ladders suggest that the stabilization of spin structures consisting of 
4 spins in a square plaquette  as a result of resonance valence bond (RVB) physics 
suppresses the hole crystal with a  1/4 wave vector. 

\end{abstract}

\maketitle

Two years after the discovery of high-T$_{C}$ cuprates \cite{Bednorz},
a new phase of Cu-O based systems, Sr$_{14-x}$Ca$_x$Cu$_{24}$O$_{41}$ (SCCO),
was found \cite{McCarron,Siegrist}. SCCO is a layered material with
two different types of copper oxide sheets -- one with CuO$_2$ 'chains' 
and one with Cu$_2$O$_3$ 'ladders' (see Fig. 1a).  The sheets are separated 
by Sr/Ca atoms, and are stacked in an alternating fashion along the b
crystallographic direction. They are structurally incommensurate; the
ratio of chain and ladder lattice parameters is $c_c$/$c_{L}$$\approx$7/10. 
From charge neutrality, the formal valence of 
Cu is $+2.25$, resulting in 6 holes per unit cell if counting from a 
Cu$^{2+}$ state. The substitution of Ca for Sr redistributes the holes between 
the chains and the ladders allowing for studies as a function of hole 
density without the influence of strong scattering by charged impurities. 

A hole-doped two-leg spin ladder is the minimum needed to obtain
superconductivity (SC) \cite{Dagotto1,Dagotto2,Siegrist}, although
this competes with an insulating 'hole crystal' (HC) phase
\cite{Dagotto1,White,Carr}. SC with $T_C\approx$12 K has indeed been
found by Uehara {\em et. al}~\cite{Uehara} for samples with $x = 13.6$
under hydrostatic pressure $>$ 3 GPa. Recently, the HC phase in the ladders 
was also discovered by resonant soft x-ay scattering (RSXS)
\cite{Abbamonte,Rusydi}. The HC is observed to have only odd
periodicity ($\lambda_{HC}=5c_L$ for $x=0$ and
$\lambda_{HC}\approx 3 c_L$ for $x=10, 11$ and 12). The competition
between the HC and SC phases in the 
two-leg spin ladder is believed to be similar to that between ordered
stripes and SC in doped two-dimensional Mott insulators
\cite{Tran1} making it an important model system to try to understand. 
In fact, a two-dimensional model of coupled two-leg spin
ladders \cite{UhrigPRL04} was used to explain recent neutron scattering
data of two-dimensional cuprates La$_{15/8}$Ba$_{1/8}$CuO$_4$ \cite{TranNature04}.

A serious problem exists however in understanding the periodicity of the 
HC in the ladders in terms of what was thought to be the correct doped hole 
density of the ladders. The often 
accepted doped hole density of $\delta_L\approx1/14$ (i.e $\sim1$ hole for every 14 Cu's)
in the ladder  and $\delta_c\approx5/10$ (i.e. $\sim5$ holes for every 10 Cu's) in the 
chain is a much too small density to 
arrive at a HC periodicity of 5$c_L$ in 
ladder units. Until we understand this we really cannot reach any 
conclusions with regard to proposed models predictions 
regarding charge and spin ordering in  these structures. 

The present paper deals with this problem and presents a new interpretation of 
the polarization dependent x-ray absorption data which leads to a much 
different charge density distribution between ladders and chains and with 
which the observed periodicities of hole crystals emerges quite naturally. 
We conclude that $\delta_L$ and $\delta_c$ are 2.8/14 and 3.2/10
for $x=0$, respectively, and the number of holes in the ladder increases 
almost linearly with $x$, reaching $\delta_L=4.4/14$ 
and $\delta_c=1.6/10$ 
for $x=11$. The end result is a strong support of the rung based hole 
pairing predicted using simple $t-J$ like models \cite{White}. We attribute the
absence of the HC with $\lambda=4c_L$, expected at $x=4$ but not
observed, to simple RVB-physics.

The distribution of holes between the chains and ladders is essential in 
determining the answer for the problem we mentioned above, and most  
physical properties. Like in high-$T_{\rm C}$ cuprates, the holes are expected 
to enter into O 2$p$ orbitals and to form spin compensated local bound states with a Cu
3$d$ hole \cite{EskesPRL88} refered to as Zhang-Rice (ZR) singlets
\cite{Zhang}.  Various experiments suggest that $\delta_L\approx1/14$ 
and $\delta_c\approx5/10$ at $x=0$, and provide evidence for a transfer of 
holes from the chains to the ladders upon Ca substitution. X-ray absorption 
measurements \cite{Nucker} find that $\delta_L$ ranges from 0.8/14 for 
$\it{x}$ = 0 to 1.1/14 for $\it{x}$ = 12. Optical conductivity data \cite{Osafune} find a
range from 1/14 to 2.8/14, while $^{63}$Cu NMR studies \cite{Magishi} find a
range from 1/14 to 3.5/14. However, the optical conductivity and the NMR
data analysis at $x=0$ are based on the neutron diffraction
observation \cite{Matsuda} of what was thought to be a superlattice
reflection corresponding to a chain charge density wave (CDW) consistent with
$\delta_c=5/10$, and thus $\delta_L=1/14$. More recent neutron studies
\cite{Etrillard,Braden} and analysis of the crystal structure by van Smaalen 
\cite{Smaalen} have clearly shown that this peak is expected
in the basic crystal structure and  is therefore not evidence for a CDW. 

\begin{figure}
\includegraphics[width=83.mm]{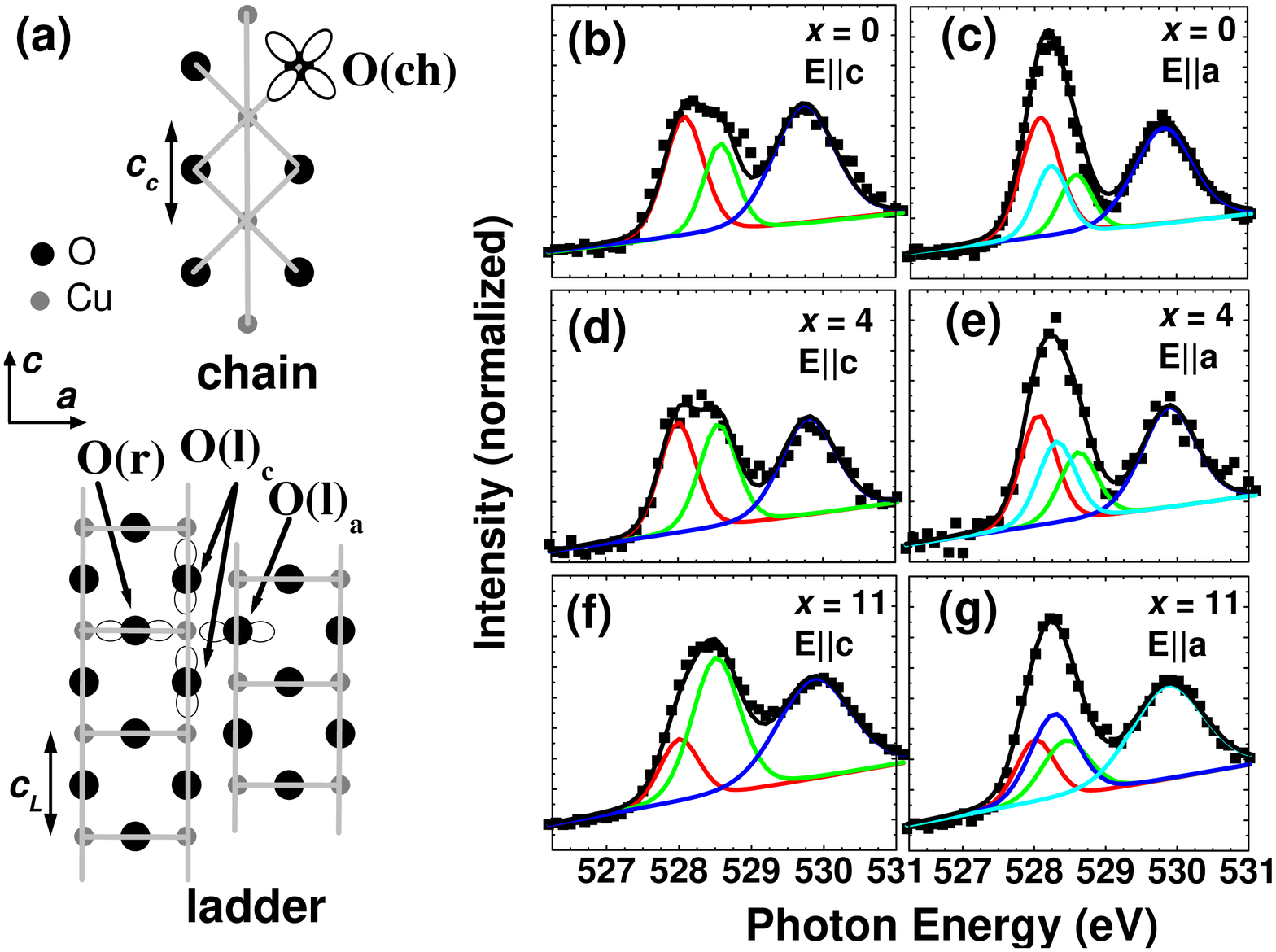}
\caption{(a) Structure of chains and ladders. The orientation of the O
2$p$ orbitals involved in the ZR singlets are indicated. The three
different oxygen sites are identified for the ladder. The right-side
panels show XAS for E$\|c$ and E$\|a$ and its theoretical fits, for
$x$ = 0 (b,c), $x$ = 4 (d,e), and $x$ = 11 (f,g). Squares are the
experimental data. Black, red, green, cyan and blue lines, are
theoretical curve for total fitting, respectively contributions of
O(ch), O(l), O(r) and UHB.  }
\end{figure}

The only direct measurement of the hole density distribution comes from
polarization dependent XAS. This is also subject to interpretation,
and the model used previously had unexplained discrepancies with
regard to the polarization-dependence.
In N\"{u}cker {\em et. al}'s analysis \cite{Nucker} of the XAS
data it is concluded that the holes are mainly concentrated on the
chains. Their interpretation assumes 
that there are only 2 distinct O 1$s$ pre-edge absorption
energies, one (H1) corresponding to holes in the chains and the other
(H2) to holes in the ladders.  The H1 peak should be independent of
the ac-plane polarization since the lobes of the O 2$p$ orbitals
involved in the chain ZR singlets are oriented at 45$^o$ to the $a$-
and $c$-axes (see Fig. 1a). The H2 peak should be strongly
polarization dependent since the rung O and the leg O have different hole
amplitudes. However, for $x=0$, XAS data shows that H1 is as
strongly polarized as H2. In Ref. \cite{Nucker} it is argued that
this effect is small and therefore is neglected.

Single crystals of SCCO were grown by travelling solvent floating zone
techniques \cite{Motoyama}.  The surfaces were prepared in the manner
described in Ref. \cite{Abbamonte}.  Polarization-dependent XAS measurements 
in the fluorescence detection mode were carried out
on the soft x-ray undulator line X1B at the National Synchrotron Light
Source. The energy resolution in the range of interest was about 200
meV. The spectra were corrected for incident flux variations and were
normalized at about 70 eV above and 10 eV below the edge where the
absorption is atomic like and structureless.

The $x=0$ spectrum shown in Fig. 1(b)\&(c) is identical to that
published in Ref. \cite{Nucker}.  Like them, we assign the lowest
energy structures to the holes doped in O 2$p$ orbitals and the higher
energy structure at about 530 eV to transitions to the upper Hubbard
band (UHB) or Cu 3$d$ orbitals. These are followed by a broader
structure due to transitions to unoccupied bands hybridized with O
2$p$ and 3$p$ states.  The UHB structure is only weakly polarization
dependent, as expected given the symmetry of the empty $d_{x^2-y^2}$
orbital.  Since the point group symmetry for the ladder is not quite
D$_{4h}$ (the four O surrounding a Cu are not identical, see Fig. 1a)
some polarization-dependence remains. In addition, there is a strongly
polarization dependent feature at lower energies, composed of at least
two components.  In Ref. \cite{Nucker}, XAS data for
La$_{3}$Sr$_{11}$Cu$_{24}$O$_{41.02}$ (with only 3 holes per unit
cell) shows only one nearly polarization independent structure at the
lower energy.  As concluded there, this strongly suggests that the
holes involved are almost solely on the chains, where all O sites O(ch)
are identical, and with 2$p$ orbitals oriented to 45$^o$ (see
Fig. 1a).

\begin{figure}
\includegraphics[width=82.mm]{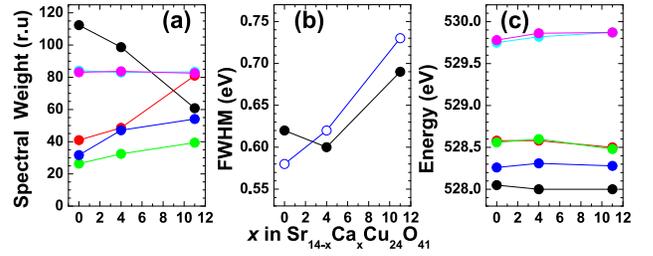}
\caption{(a) Spectral weight (SW) (b) full width at half maximum
(FWHM) and (c) energy of O(ch) (black), O(l)$_c$ (red), O(l)$_a$
(green), O(r) (blue), UHB for E$\|c$ (cyan) and UHB for E$\|a$
(magenta) as function of $x$. Blue open-circles are for oxygen ions in
the ladders.  }
\end{figure}

In ladders, things are more complicated. There are two types of O
sites: the rung sites, O(r), coordinated by 2 Cu ions, and the leg
sites, O(l), coordinated by 3 Cu ions (see Fig. 1(a)). The different
coordination numbers result in different binding energies for the core
1$s$ and valence 2$p$ orbitals. Higher values are expected for the
orbitals of O(l), while those of O(r) should be close to O(ch) which
is also coordinated by 2 Cu ions. Moreover, each ladder ZR singlet
involves one O(r)$_a$, two O(l)$_c$ from the leg, and one O(l)$_a$
from the leg of a neighboring ladder. The subscripts $a$ and $c$ refer
to the polarization needed to observe O 1$s$ to 2$p$ transitions. For
$a$-polarization, transitions are possible for O(r)$_a$ and O(l)$_a$
at different energies, while for $c$-polarization, transitions are
only possible from 2 identical O(l)$_c$, with energy close to that of
O(l)$_a$.

We performed a simultaneous least-square fit to all the measured
spectra. Outputs of the fits, namely the spectral weight (SW), full
width at half maximum (FWHM) and various energies are shown in
Fig. 2(a)-(c). The number of holes $\delta_L=(3-5\delta_c)/7$ are determined from the
spectral weights of the various absorption lines:
\begin{equation}
\label{eq1}
\delta_L = 
\frac{3(SW_{O(r)_a}+SW_{O(l)_a}+SW_{O(l)_c})}
{7(SW_{O(r)_a}+SW_{O(l)_a}+SW_{O(l)_c}+SW_{O(ch)})}
\end{equation}
where $SW_{O(ch)}$ is the total SW for both polarizations.

We start with $x=0$. XAS data and theoretical fits are shown in
Figs. 1(b),1(c). For E$\|c$, the doped hole region has at least two
structures. In our interpretation this is due to the energy difference
between O(ch) and O(l)$_c$.  For E$\|a$, the contribution of O(l)$_c$
should be replaced with that of O(l)$_a$ and O(r)$_a$, thus shifting
more weight to the lower energy.  This is indeed consistent with the
data. In our fitting results, shown in Fig. 2, the energy of O(r)$_a$
is about 0.2 eV higher than that of O(ch), while the energy of
O(l)$_c$ and O(l)$_a$ are roughly equal and about 0.5 eV higher than
that of O(ch). The SW of the UHB is almost polarization
independent. This shows that the ladder holes are distributed nearly
isotropically amongst the 4 O involved in the ZR singlet, even though
the symmetry is not the full D$_{4h}$.  (If large deviations were
found, the ZR picture would not be valid for the ladder holes).  The
FWHM of O(ch), which is about 0.62 eV, is about 5\% larger than the
ones in the ladders.  From Eq. (1), we find $\delta_L=2.8/14$ and
$\delta_c=3.2/10$.

We continue the analysis for $x=4$. In Figs. 1(d),1(e) we show the
fitting for E$\|c$ and E$\|a$ data, respectively. Again, the SW in the
UHB is almost polarization independent (Fig. 2(a)) while the energies
of the various O sites are close to the $x=0$ values (Fig. 2(c)),
consistent with our understanding of the polarization dependent
XAS. From Eq. (1) we find $\delta_L=3.4/14$ and $\delta_c=2.6/10$.

Figs. 1(f) and  1(g) show our XAS data for $x=11$. (The maximum HC intensity 
occurs at $x=11$ where the wave vector is closest to 3c$_L$ \cite{Rusydi}). 
In Ref. \cite{Nucker} it is claimed that here, the UHB is strongly 
polarization dependent while the O doped hole pre-edge region is 
almost polarization independent. Our results show an opposite behavior, 
like for samples with  smaller $x$. In fact, the shape and the intensity of the hole-doped
and UHB peaks of Ref. \cite{Nucker} are similar to ours, and can be
made to coincide by rescaling.  We conclude that the difference is due
to the method used in the normalization of the data.

The fit of the polarization dependent XAS for $x=11$ is more
difficult than for $x=0$ or 4.  The problem is not statistical noise,
but rather the structural peaks themselves. We see in Fig. 1(f) \& (g)
that there is no clear evidence for multiple peaks in the doped hole
regime because the energy resolution is too poor to resolve the peaks.
We therefore use the $x=0$ results as input for the
fitting. The fits are shown in Figs. 1(f), 1(g). As before, we find
that the SW of UHB is almost polarization independent and the energies
of the various O sites remain close to the $x=0$ values (see Fig 2),
validating our interpretation. We find $\delta_L=4.4/14$ and 
$\delta_c=1.6/10$. It is important to mention that these results depend 
strongly on the energy of O(ch). For example, varying it by 0.05 eV 
changes $\delta_L$ by about 0.5/14. This is due to an instability in the fitting
process caused by the close proximity of peaks in the pre-edge region
(the energy of O(ch) falls close to the leading edge of the hole-doped
peak). This is why the $x=0$ spectrum provides an important reference.

\begin{figure}
\includegraphics[width=75.mm]{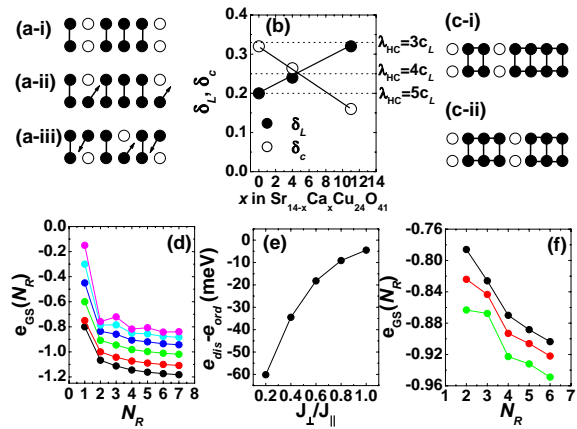}
\caption{(a-i)-(a-iii) Scenarios for the ladder hole distribution in a HC
  with $\lambda=4c_L$, as predicted by DMRG. Filled (open) circles
  represent electrons (holes). Arrows indicate the spin of unpaired
  electrons.
  (b) The estimated number of
  holes in the ladders (filled circles) and the chain (open
  circles). Dashed lines show the number of holes for scenario (a-i),
  for $\lambda_{HC}/c_L$= 3, 4, and 5. 
  (c-i) Disordered and (c-ii) ordered hole arrangement corresponding to
  $\lambda_{HC}= 4c_L$. The
  disordered state has equal numbers of 2-rung and 4-rung spin
  plaquettes. (d) GS energy per rung, in units of $J_\parallel$, as a
  function of the number of rungs in the spin plaquette.  Black, red,
  green, blue, cyan, and magenta-filled circles are results for
  $J_\perp$/$J_\parallel$ = 1.2, 1.0, 0.8, 0.6, 0.4 and 0.2,
  respectively. (e) $e_{dis} - e_{ord}$ as a function of $J_\perp$/$J_\parallel$, for
  $J_\parallel$ = 130 meV (see text for details). (f) $e_{GS}(N_R)$
  vs. $N_R$  for 
  $J_\perp$/$J_\parallel$ = 0.4 and $J_{ring}$ /$J_\parallel$ = -0.1,
  0, and 0.1 (black, red and green-filled circles, respectively). 
}

\end{figure}

We now analyze some possible scenarios of the hole distribution in the
ladder, for these new $n_L$ values. We also consider the connection to
the wavelengths $\lambda_{HC}=3c_L$ ($x=11$) and 5$c_L$ ($x=0$)
of the recently discovered HC~\cite{Abbamonte, Rusydi}. First, DMRG
calculations for a single ladder \cite{White} found that holes prefer to
pair along the rungs, resulting in a charge density wave shown
pictorially in Fig. 3 (a-i).  For periodicity $\lambda_{HC}=N c_L$,
the doped hole density in ladder $\delta_L$ should be 1/N, i.e. $\delta_L=0.2$ 
if $N=5$, $\delta_L=0.25$ if $N=4$ and $\delta_L=0.333$ if $N=3$.  These values are very
close to our new XAS estimates, as shown in Fig. 3(b). Surprisingly,
however, the $\lambda_{HC}=4c_L$ HC, which is predicted by DMRG
\cite{White,Carr} to be a stable phase, is not observed in RSXS
\cite{Rusydi}. While the study of this discrepancy is on
going, a possible explanation is proposed below. Another hole
distribution consistent with these $\lambda_{HC}$ values, not
involving pairing, and also discussed in Ref. \cite{White}, is single rung 
bond-centered holes which is shown in Fig. 3(a-ii). 
Since $\delta_L$ for this case is halved, this model does not
match our XAS results. In a third scenario proposed in
Ref. \cite{White}, the holes could be site centered, alternating
between the two legs as shown in Fig. 3(a-iii).  While this matches the
$\delta_L$ values, it requires an odd number of undoped rungs in the HC
unit cell and is therefore inconsistent with the observed $N=3$ and
$N=5$ HC. We conclude that our results support the scenario of holes
paired along the rungs.

Assuming rung-paired holes, a minimal model of the doped ladder is an
antiferromagnetic Heisenberg Hamiltonian plus a cyclic four spin
exchange term \cite{Takahashi1977,Brehmer}:
\begin{equation}
\nonumber {\cal H} = J_{\parallel} \sum_{n=1\atop
\alpha=1,2}^{N_R-1}{\bf S}_{\alpha,n}\cdot {\bf S}_{\alpha,n+1} +
J_{\perp} \sum_{n=1}^{N_R}{\bf S}_{1,n}\cdot {\bf S}_{2,n} + {\cal
H}_{\rm ring}
\end{equation}
Here, $N_R=N-1$ is the number of undoped rungs per HC unit cell,
$\alpha=$ 1,2 indexes spins on the two legs, and $J_\perp$ and
$J_\parallel$ are exchange couplings along the rung and leg,
respectively. We assume no coupling between spins on opposite sides of
rungs occupied by paired holes.  ${\cal H}_{\rm ring}=J_{\rm ring}
\sum_{n=1}^{N_R-1} [{1\over 4} + {\bf S}_{1,n}\cdot{\bf
S}_{1,n+1}+{\bf S}_{2,n}\cdot{\bf S}_{2,n+1} +{\bf S}_{1,n}\cdot{\bf
S}_{2,n+1}+{\bf S}_{2,n}\cdot{\bf S}_{1,n+1}+{\bf S}_{1,n}\cdot{\bf
S}_{2,n} +{\bf S}_{1,n+1}\cdot{\bf S}_{2,n+1}+4\{ ({\bf
S}_{1,n}\cdot{\bf S}_{2,n})({\bf S}_{1,n+1}\cdot{\bf S}_{2,n+1}) +
({\bf S}_{1,n}\cdot{\bf S}_{1,n+1})({\bf S}_{2,n}\cdot{\bf S}_{2,n+1})
-({\bf S}_{1,n}\cdot{\bf S}_{2,n+1})({\bf S}_{2,n}\cdot{\bf
S}_{1,n+1})\} ]$ is a four-spin cyclic exchange.

We use exact diagonalization to find the ground state for various
$N_R$ values. The ground-state (GS) energy per rung, $e_{GS}(N_R)=
E_{GS}(N_R)/N_R$, is shown in Fig. 3 (d) for various ratios of
$J_\perp$/$J_\parallel$ and J$_{\rm ring}$=0. The value of
$J_\perp$/$J_\parallel$ is not known accurately, but is believed to be
between 0.5 and 1.13 \cite{Imai,Magishi,Kumagai,Carretta,Takigawa,Gozar}. 
An even/odd oscillation is observed for small $N_R$ and $J_\perp$/$J_\parallel$ $<$ 1,
favoring $N_R=2$ and $4$ ($\lambda_{HC}=3, 5c_L$). The origin of this
oscillation is simple. The limit $J_{\parallel}\gg J_{\perp}$
corresponds to two AFM chains weakly coupled along the rungs. For even
$N_R$, spins on each leg pair in a RVB-like state, and $E_{GS}$ is
low. For odd $N_R$, each leg has an unpaired spin, significantly
increasing $E_{GS}$. In the limit $J_{\perp}\gg J_{\parallel}$, the GS
consists of spin-singlets along the rungs and the parity of $N_R$ is
irrelevant.  At large $N_R$, $e_{GS}$ converges to the bulk value. This
even/odd oscillation provides a possible explanation for the absence
of a HC with $N_R=3$ ($\lambda_{HC}=$ 4$c_L$).  This HC costs an
energy $e_{\rm ord}=6e_{GS}(3)$ per two unit cells (see Fig. 3(c-ii)). 
A disordered phase, at the same doping, has equal numbers of $N_R=2$ 
and $N_R=4$ plaquettes and an energy $e_{\rm dis}=2 e_{GS}(2)+4e_{GS}(4)$ 
for the same length (see Fig. 3(c-i)). If $e_{dis} < e_{ord}$, the HC 
phase is unstable. For $J_\parallel$ = 130 meV, we plot $e_{dis} - e_{ord}$ 
in Fig. 3(e), showing that the disordered phase is energetically favorable,
especially for lower values of $J_\perp$/$J_\parallel$. 

We have also studied the effect of ${\cal H}_{\rm ring}$ on
$E_{GS}$. Such terms appear in $4^{th}$ order perturbation expansions
in the strong coupling limit of the Hubbard model \cite{Takahashi1977}
and are known to play an important role for Wigner crystals and $^3$He
solid. Typical results for $e_{GS}(N_R)$ are shown in Fig. 3(f), for
$J_\perp$/$J_\parallel$ =0.56 \cite{Eccleston} and 
$J_{\rm ring}$/$J_{\parallel} =-$0.1, 0, and 0.1. Since the sign of 
the ring exchange and superexchange should be the same \cite{Takahashi1977}, 
it follows that a large J$_{\rm ring}$ suppresses the even-odd effect. For example, for
$J_\perp$/$J_\parallel$ =0.56, $e_{dis} - e_{ord}$ increases from
-20.8meV if $J_{\rm ring}=0$, to -12.5 meV if $J_{\rm
  ring}$/$J_{\parallel} =$0.1. We conclude that for reasonable values
of $J_\perp$, $J_{\parallel}$ and $J_{\rm ring}$  this simple model
offers a possible explanation for the absence of the
$\lambda_{HC}\sim$ 4$c_L$ HC. An accurate determination of the
exchange couplings is needed before the issue can be settled.

In conclusion, we propose a new interpretation of polarization
dependent XAS for SCCO. Based on our analysis combining the XAS and 
RSXS data, we find strong support for a pairing of holes in the rungs of 
the ladders. We also give a possible explanation for the absence
of the HC with 1/4 periodicity in terms of RVB physics.

We acknowledge helpful discussions with I. Affleck, A. Sandvik, G. Blumberg 
and J. Zaanen. This work was supported by FOM, US Department of Energy,  
Canadian funding agencies: NSERC, CIAR, and CFI, the 21st Century COE program of the 
Japan Society for Promotion of Science, and
the Helmholtz Association contract VH-FZ-007.


\end{document}